\documentclass{aip-cp}

\usepackage[numbers]{natbib}
\usepackage{rotating}
\usepackage{graphicx}

\begin{document}

% Title portion
\title{Constraining the neutron star equation of state using Pulse Profile Modeling}

\author[aff1]{Anna L. Watts\corref{cor1}}
%\eaddress[url]{http://www.aip.org}

\affil[aff1]{Anton Pannekoek Institute, University of Amsterdam, PO Box 94249, 1090 GE Amsterdam, The Netherlands}

\corresp[cor1]{Corresponding author: A.L.Watts@uva.nl}

\maketitle

\begin{abstract}
One very promising technique for measuring the dense matter Equation of State exploits hotspots that form on the neutron star surface due to the pulsar mechanism, accretion streams, or during thermonuclear explosions in the neutron star ocean. This article explains how {\it Pulse Profile Modeling} of hotspots is being used by the Neutron Star Interior Composition Explorer (NICER), an X-ray telescope installed on the International Space Station in 2017 - and why the technique is a mission driver for the next, larger-area generation of telescopes including the enhanced X-ray Timing and Polarimetry (eXTP) mission and the Spectroscopic Time-Resolving Observatory for Broadband Energy X-rays (STROBE-X).
\end{abstract}

% Head 1
\section{INTRODUCTION}

\begin{figure}[h]
  \centerline{\includegraphics[width=450pt]{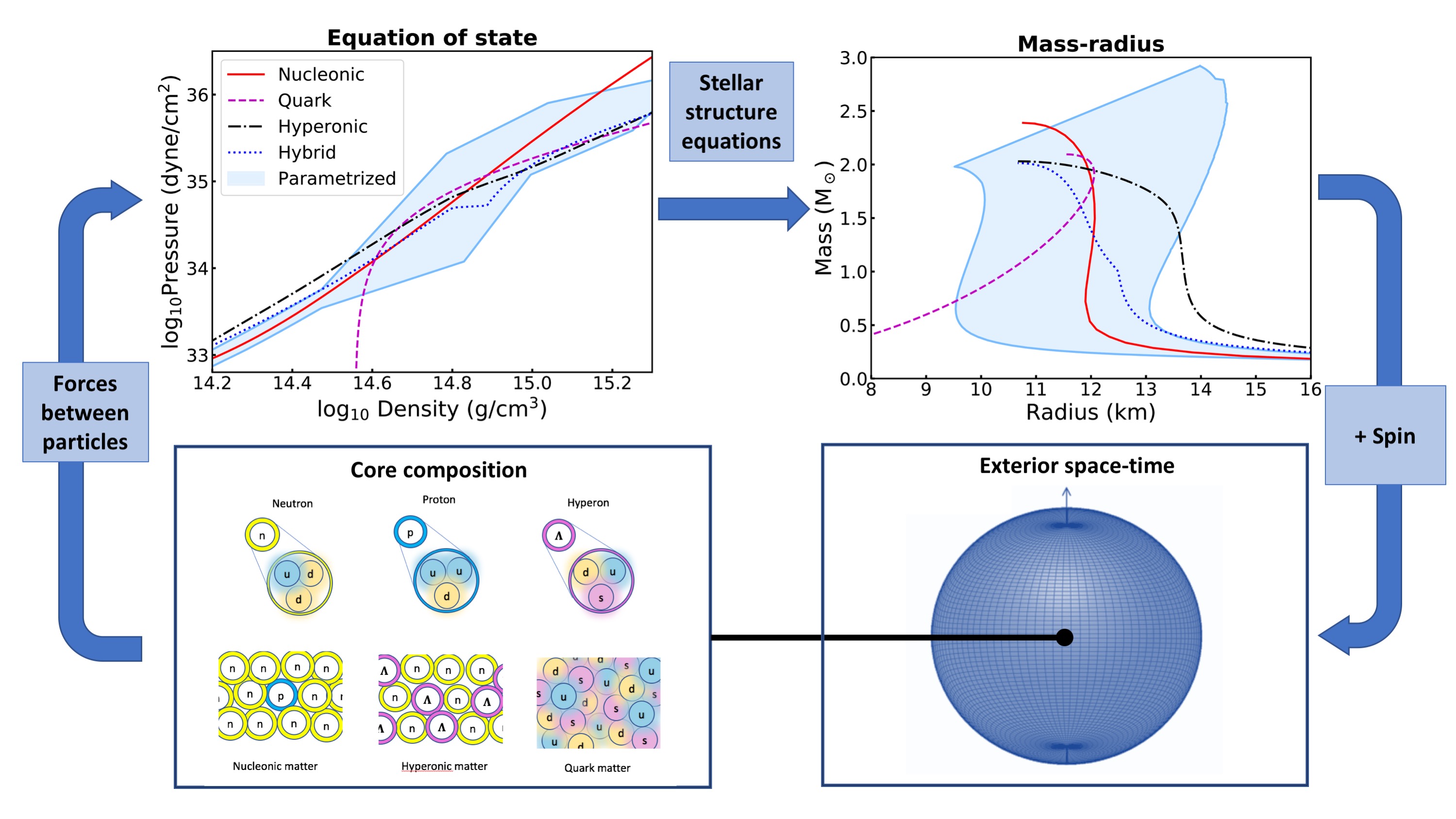}}
  \caption{The relationship between the composition and inter-particle forces in the neutron star core, the EOS, the mass-radius relation, and the exterior space-time of the star. The space-time of the rotating neutron star imprints its signature on radiation emitted from the stellar surface: we can use this to infer the EOS.\label{eosloop}}
\end{figure}

Densities in the cores of neutron stars can reach up to ten times that of normal nuclear matter. In addition to nucleonic matter in conditions of extreme neutron-richness, neutron stars may also contain stable states of strange matter; either bound in the form of hyperons or in the form of deconfined quarks. The extremes of the Quantum Chromodynamics phase diagram cannot currently be explored using first principles calculations, due to the numerical challenges involved. We rely instead on phenomenological models of particle interactions and phase transitions, which are tested by experiment and observation. Heavy ion collision experiments explore the high temperature and lower density parts of the phase diagram; but neutron stars are unique laboratories for the study of strong and weak force physics in cold, ultra-dense matter \citep[for recent reviews see][]{Lattimer16,Oertel17,Baym18}. Our uncertainties about the microphysics of particle interactions in the conditions that prevail inside neutron stars are codified in the Equation of State (EOS), the relation between pressure and (energy) density. The EOS forms part of the relativistic stellar structure equations that enable us, given a central density and a spin rate, to compute model neutron stars. The EOS parameters are mapped, by these equations, to parameters such as mass, radius or tidal deformability that determine the exterior space-time of the star (Figure \ref{eosloop}). So if we can measure mass and radius - for a range of stellar masses - we can in principle map out the EOS.

Mass can be measured directly via pulsar timing, and if sufficiently high can rule out EOS models: the discovery of 2M$_\odot$ neutron stars \citep{Demorest10,Antoniadis13} has posed big challenges for our understanding of hyperons \citep{Chatterjee16}. Obtaining radius via pulsar timing is harder; one good measurement is expected with the Square Kilometer Array \citep{Watts15}, but more will be needed to map the EOS. We must therefore turn to other techniques. Mass and tidal deformability can in principle be measured by gravitational wave  observations of binary neutron star mergers. The detection of GW170817 and its unusual electromagnetic counterpart have yielded some results \citep[e.g.][]{Abbott18,Annala18} but tens of detections are likely to be needed to map the EOS \citep{Agathos15} and there are still uncertainties in the template models \citep{Lackey15}. One can also use the fact that the exterior space-time affects X-rays emitted from the stellar surface \citep[see][for a review]{Ozel13}. Spectral modeling of bursting or quiescent neutron stars is one option \citep[e.g.][]{Nattila17,Shaw18}, however issues like compositional and distance uncertainty affect reliability \citep{Miller13}. Here I focus on an alternative technique, Pulse Profile Modeling \citep[see][for a review]{Watts16}. Pulse Profile Modeling is being used by NASA's Neutron Star Interior Composition Explorer \citep[NICER,][]{Gendreau16}, a pioneering soft X-ray telescope installed on the International Space Station in 2017. Pulse Profile Modeling is exciting because it can in principle deliver simultaneous measurement of mass and radius at an unprecedented level of a few percent \citep{Lo13,Psaltis14b}.

\section{PULSE PROFILE MODELING}

Pulse Profile Modeling (also known as waveform or lightcurve modeling) exploits the effects of General and Special Relativity on rotationally-modulated emission from neutron star surface hot spots (see Figures 4 and 5 of \cite{Watts19} for examples that illustrate these effects).  A body of work extending over the last few decades has established how to model the relevant aspects - which include gravitational light-bending, Doppler boosting, aberration, time delays and  the effects of rotationally-induced stellar oblateness - with a very high degree of accuracy \citep{Pechenick83,Miller98,Poutanen03,Poutanen06,Cadeau07,Morsink07,Baubock13,AlGendy14,Psaltis14a,Nattila18}.  Given a model for the surface emission (surface temperature pattern, atmospheric beaming function, observer inclination) we can thus predict the observed pulse profile (counts per rotational-phase bin per energy channel) for a given exterior neutron star space-time (set by mass, radius and spin frequency - see the review by \cite{Watts16} for a more extended introduction to Pulse Profile Modeling).  By coupling such lightcurve models to a sampler, we can use Bayesian inference to derive posterior probability distributions for mass and radius, or the EOS parameters, directly from pulse profile data.

Successful application of the Pulse Profile Modeling technique requires sources with a rapid spin ($>$100 Hz), to ensure that Special Relativistic effects are strong enough. It also requires high quality phase- and energy-resolved waveforms: time resolution $\le 10\mu$s, and a minimum number of photons.  The precise number needed to deliver constraints on mass and radius at levels of a few percent, and hence provide tight limits on EOS models, depends on the geometry of a given source - but is roughly $\sim 10^6$ pulsed photons \citep{Lo13,Psaltis14b}. The attraction of Pulse Profile Modeling is that this is not only feasible in reasonable observation times, but can also be done for three different source classes with surface hotspots: rotation-powered pulsars, accretion-powered pulsars, and thermonuclear burst oscillation sources. Each class has multiple instances, increasing the odds of sampling a wide range of masses and hence mapping more completely the EOS.

\begin{figure}[h]
  \centerline{\includegraphics[width=450pt]{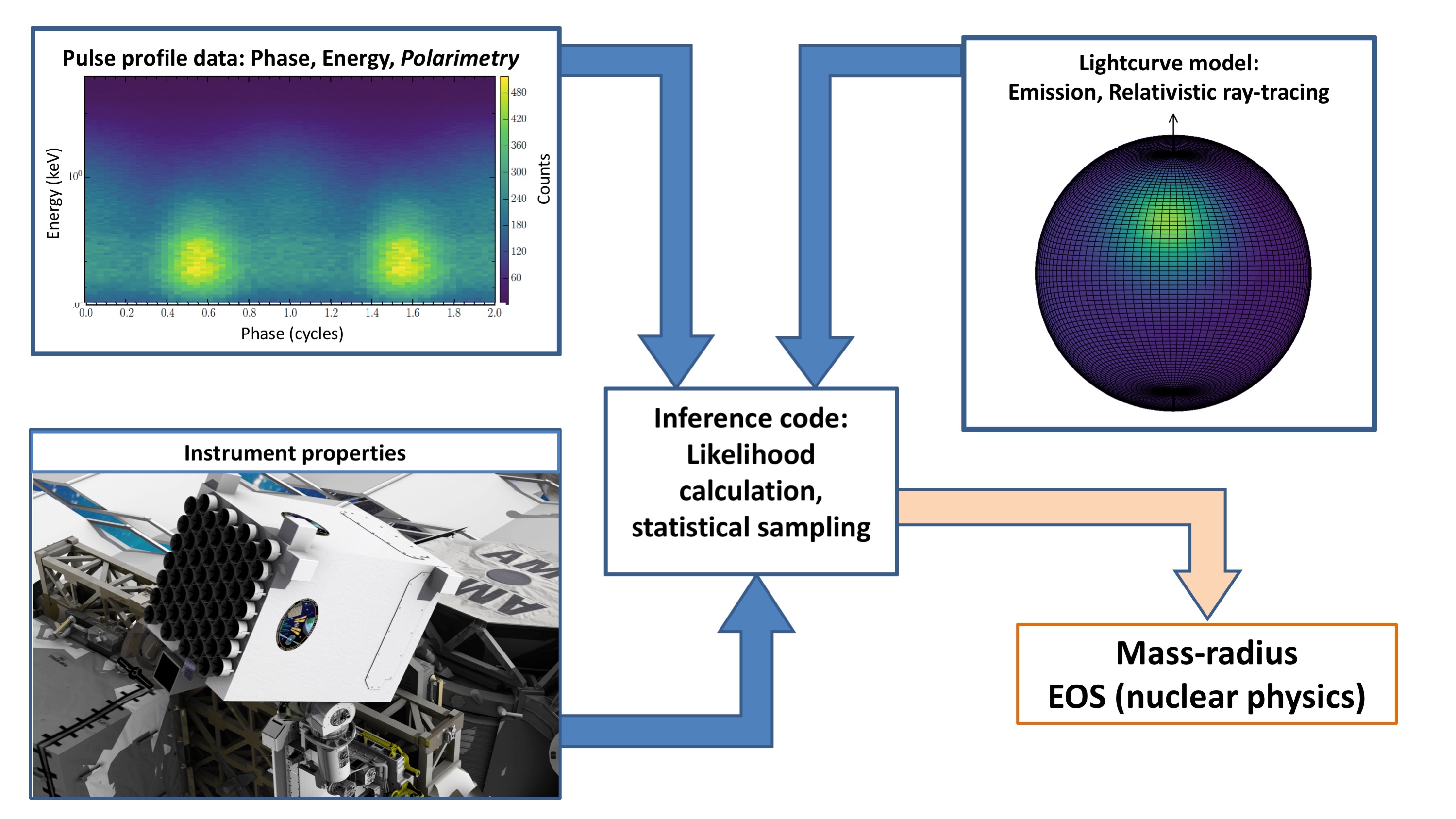}}
  \caption{The Bayesian inference process at the heart of the Pulse Profile Modeling technique. Pulse Profile Data consist of X-ray counts binned by energy and rotational phase, assuming some spin or frequency-evolution model: in the future they will also include polarimetry information. The Lightcurve Model generates a synthetic pulse profile for a given set of space-time parameters: it must include the surface radiation pattern, any beaming, observer inclination, relativistic ray-tracing and background emission.  The instrument properties must also be factored in (Image Credit for NICER telescope graphic: NASA).  Inference Codes then apply Bayes' theorem to the data for the given model, generating posterior probability distributions for either mass and radius, or EOS model parameters.\label{infprocess}}
\end{figure}

Rotation-powered pulsar hotspots arise as return currents in the pulsar magnetosphere deposit energy in the neutron star surface layers; the resulting surface temperature and beaming pattern is highly uncertain \citep{Timokhin13,Gralla17}. Rotation-powered pulsar pulse profiles are however extremely stable.  In accretion-powered pulsars \citep{Patruno12}, where accreting material is channeled towards the magnetic poles of the star, the pulsed emission has two main components: one from hotspots at the polar caps where the accreting material impacts the star, and one from the shock in the accretion funnel \citep{Poutanen06}. A third pulsed component may arise due to reflection from the accretion disk \citep{Wilkinson11}. Temperature pattern and beaming function are a priori uncertain. Accretion-powered pulsars also have pulse profile variations as the accretion flow changes, on timescales of days \citep{Hartman08}.  Thermonuclear burst oscillations are hotspots that form during thermonuclear (Type I X-ray) bursts in the oceans of accreting neutron stars  \citep{Galloway08,Watts12}, the bursts themselves being caused by unstable burning of accreted hydrogen, helium, or carbon. Burst emission has a well understood beaming function due to the sub-surface thermal origin \citep{Suleimanov11}, but the surface temperature pattern for thermonuclear burst oscillations remains uncertain since the mechanism responsible for generating them is not yet clear. Thermonuclear bursts are relatively short ($\sim$ 10-100s), so data from several bursts must be combined to accumulate sufficient photons for successful Pulse Profile Modeling; and the thermonuclear burst oscillation frequency and amplitude are often variable during a burst, implying that hotspot properties (size, temperature distribution, and location) are changing on short ($\sim$ 1s) timescales.    The inference process, summarized in Figure \ref{infprocess}, thus involves the following elements:

\begin{enumerate}
\item Pulse profile data: X-ray counts binned by energy and rotational phase; for rotation-powered pulsars and accretion-powered pulsars the pulsar spin ephemeris is used to generate the phase-model.  Thermonuclear burst oscillations can have drifting frequencies and this must also be taken into account.
\item A Lightcurve Model:  This generates a synthetic pulse profile for a given set of parameters including those setting the exterior space-time (mass, radius, spin rate), the surface emission pattern (temperature distribution, beaming function or atmosphere model), observer inclination, distance, interstellar hydrogen column density, and a background model.  
\item The Instrument Response Function: This determines how a specific detector would process the model incident signal.  The Redistribution Matrix File and Auxiliary Response File will map, for example, from energy space into detector channels, and take into account the overall efficiency of the detector.  
\item An Inference Code: Bayesian inference codes couple likelihood evaluations (evaluating the probability of obtaining the data from the model for a given set of parameters) and a set of priors for those parameters, with a sampler that explores the parameter space to determine the posterior probability distributions of the various model parameters.  
\item An EOS model:  If one wants to infer the EOS (rather than just mass and radius), one also needs either a set of models with fixed EOS parameters for model comparison, or a parametrized EOS model with priors for those parameters. The EOS model can be incorporated directly into the Lightcurve Model; alternatively EOS inference can be performed separately using mass-radius posteriors, subject to certain restrictions (see the later discussion on EOS constraints).
\end{enumerate}

\section{PULSE PROFILE MODELING MISSIONS}

\begin{figure}[h]
  \centerline{\includegraphics[width=450pt]{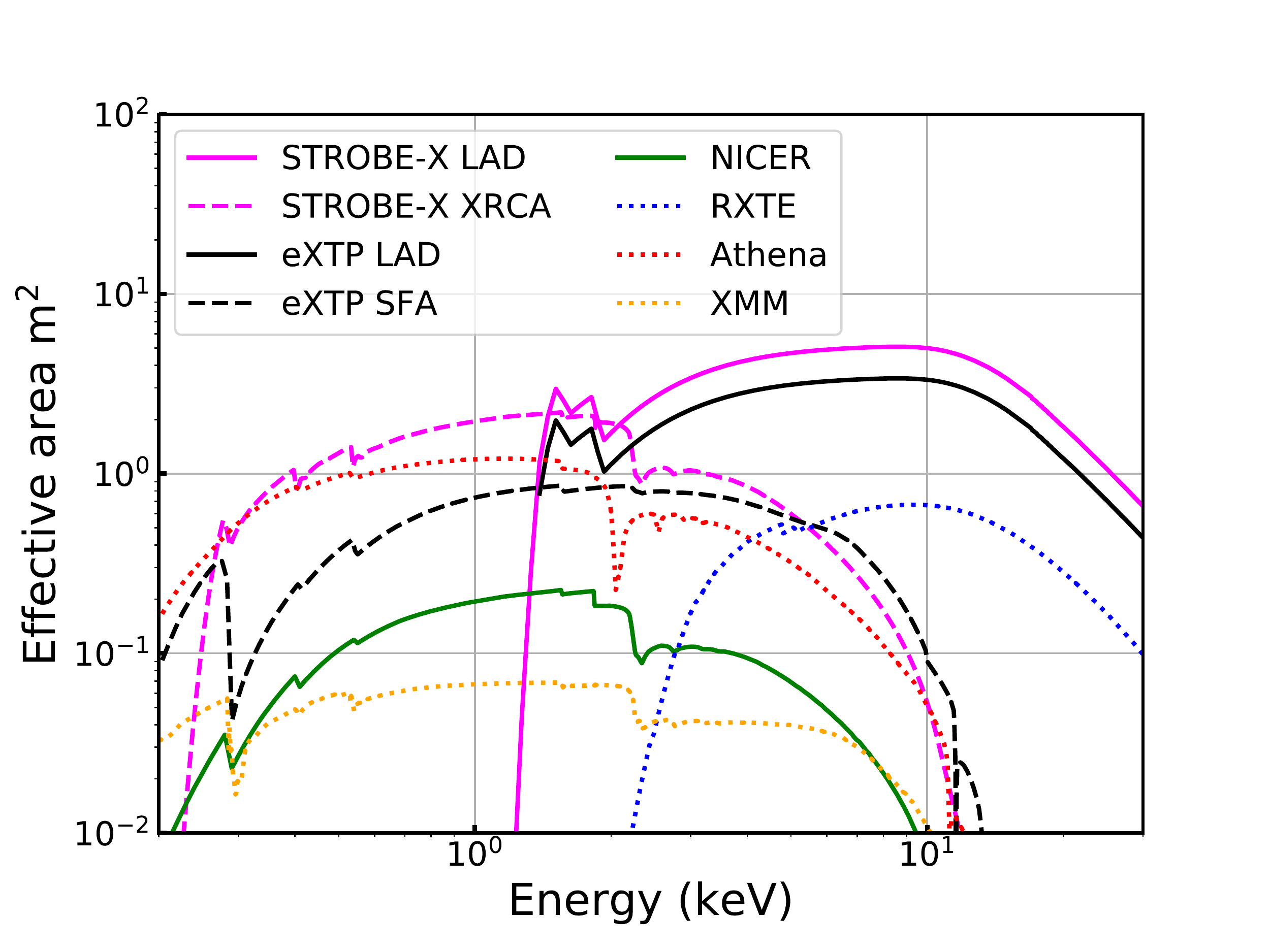}}
  \caption{Effective area curves for NICER (operational), and the proposed future large-area missions eXTP and STROBE-X, compared to other X-ray telescopes. Athena is not a timing mission, but is included for reference.\label{ppmmissions}}
\end{figure}

Preliminary attempts at Pulse Profile Modeling were made using data from the Rossi X-ray Timing Explorer (RXTE, operational from 1995-2012) and XMM-Newton (currently operational): for rotation-powered pulsars \citep{Bogdanov07,Bogdanov09,Bogdanov13}; accretion-powered pulsars \citep{Poutanen03,Leahy04,Leahy09,Morsink11}; and thermonuclear burst oscillations \citep{Bhattacharyya05}.  The uncertainties on mass and radius were large (due to low photon numbers), and these analyses did not employ the full Bayesian inference framework now in use. However they sparked interest in the technique, paving the way for the next generation of X-ray timing missions (Figure \ref{ppmmissions}).

NICER is a NASA soft X-ray telescope that was installed on the International Space Station in 2017.  NICER's primary scientific goal is to infer neutron star mass and radius using Pulse Profile Modeling for rotation-powered pulsars, where the pulse is strongest in the (0.3-1) keV range.  The pulse profile stability makes it easy to build up a good folded pulse profile using multiple exposures, even if they are well-separated in time. NICER is on track to deliver mass and radius at the 5-10\% level for four relatively bright sources (PSR J0437-4715, PSR J0030+0451, PSR J1231+1411 and PSR J2124-3358). NICER uses lightweight X-ray concentrator optics and small Silicon Drift Detectors.

Pulsations from accretion-powered pulsars and thermonuclear burst oscillation sources have a harder spectrum ($\sim$ 1-30 keV), necessitating collimating optics. Accumulating the requisite number of photons in a sensible observing time also requires a telescope with an effective area of several square meters.  For accretion-powered pulsars we expect to need observations of total duration $\sim 100$ kiloseconds to collect sufficient photons \citep{Watts19}. For thermonuclear burst oscillation sources, data from several individual bursts will need to be combined even with a large-area telescope. The observing time necessary to accumulate sufficient thermonuclear burst oscillation photons for few \% constraints on mass and radius can be estimated from the burst properties observed by RXTE over its lifetime (burst recurrence times, burst fluxes, the percentage of bursts that show thermonuclear burst oscillations, and typical thermonuclear burst oscillation amplitudes).  Estimated observation times are of order a few hundred kiloseconds \citep{Watts16}. There are currently two mission concepts for a large-area X-ray timing telescope that would access a larger, fainter population of rotation-powered pulsars than we can observe with NICER, and allow Pulse Profile Modeling with accretion-powered pulsars and thermonuclear burst oscillations: the enhanced X-ray Timing and Polarimetry (eXTP) mission, and the Spectroscopic Time-Resolving Observatory for Broadband Energy X-rays (STROBE-X). 

eXTP is a mission concept proposed by an international consortium led by the Institute of High-Energy Physics of the Chinese Academy of Sciences, with anticipated launch in 2025 \citep{Zhang19}.  The scientific payload of eXTP would consist of four instruments: the Spectroscopic Focusing Array, the Large Area Detector, the Polarimetry Focusing Array and the Wide Field Monitor.  Collectively the Spectroscopic Focusing Array (0.5-20 keV) and the collimated Large Area Detector (which uses non-imaging Silicon Drift Detectors sensitive in the 2-30 keV band) would reach a total effective area $\sim 4$ m$^2$ (Figure \ref{ppmmissions}).  The Polarimetry Focusing Array (not shown in the figure) uses Gas Pixel Detectors sensitive in the 2-10 keV range, and would have an effective area four times larger than NASA's planned Imaging X-ray Polarimetry Explorer  \citep[IXPE,][]{Weisskopf16}. The Wide Field Monitor, vital to trigger pointed observations since many accreting compact objects (including most accretion-powered pulsars and thermonuclear burst oscillation sources) are transient, will cover $\sim 4$ steradian of the sky with unprecedented sensitivity.  eXTP is about to enter Phase B (Preliminary Definition).  

In preparation for the 2020 Astrophysics Decadal Survey, NASA solicited proposals for mission concept studies for Astrophysics Probes (total lifecycle mission cost \$400 million - \$1 billion).  STROBE-X \citep{Ray18} was one of 8 proposals selected for a full 18-month concept study (the study report is now publicly available, see \cite{Ray19}).  The STROBE-X instrument suite comprises both a Wide Field Monitor and two narrow-field instruments: a soft (0.2-12keV) X-ray Concentrator Array based on NICER technology but scaled up to take advantage of a longer focal length; and a Large Area Detector using large collimated Silicon Drift Detectors.  As part of the concept study phase, the team has created detailed designs and prepared thorough cost estimates at the NASA/GSFC Instrument Design Lab.  If the mission goes forward, it would enter Phase A (Feasibility) in 2023, with an anticipated launch in 2031.

The larger area offered by eXTP and STROBE-X would allow us to make tighter inferences (since we can collect more photons) and access more sources, mapping the EOS more completely. We will be able to cross-check inference across all three source classes (rotation-powered pulsars, accretion-powered pulsars and thermonuclear burst oscillation sources); and since there are neutron stars with both accretion-powered pulsations and thermonuclear burst oscillations, we would be able to cross-check results from two types of Pulse Profile Modeling for the same source.

\section{PULSE PROFILE MODELING IN THE NICER ERA}

\subsection{MASS-RADIUS CONSTRAINTS}

\begin{figure}[!h]
\includegraphics[width=200pt]{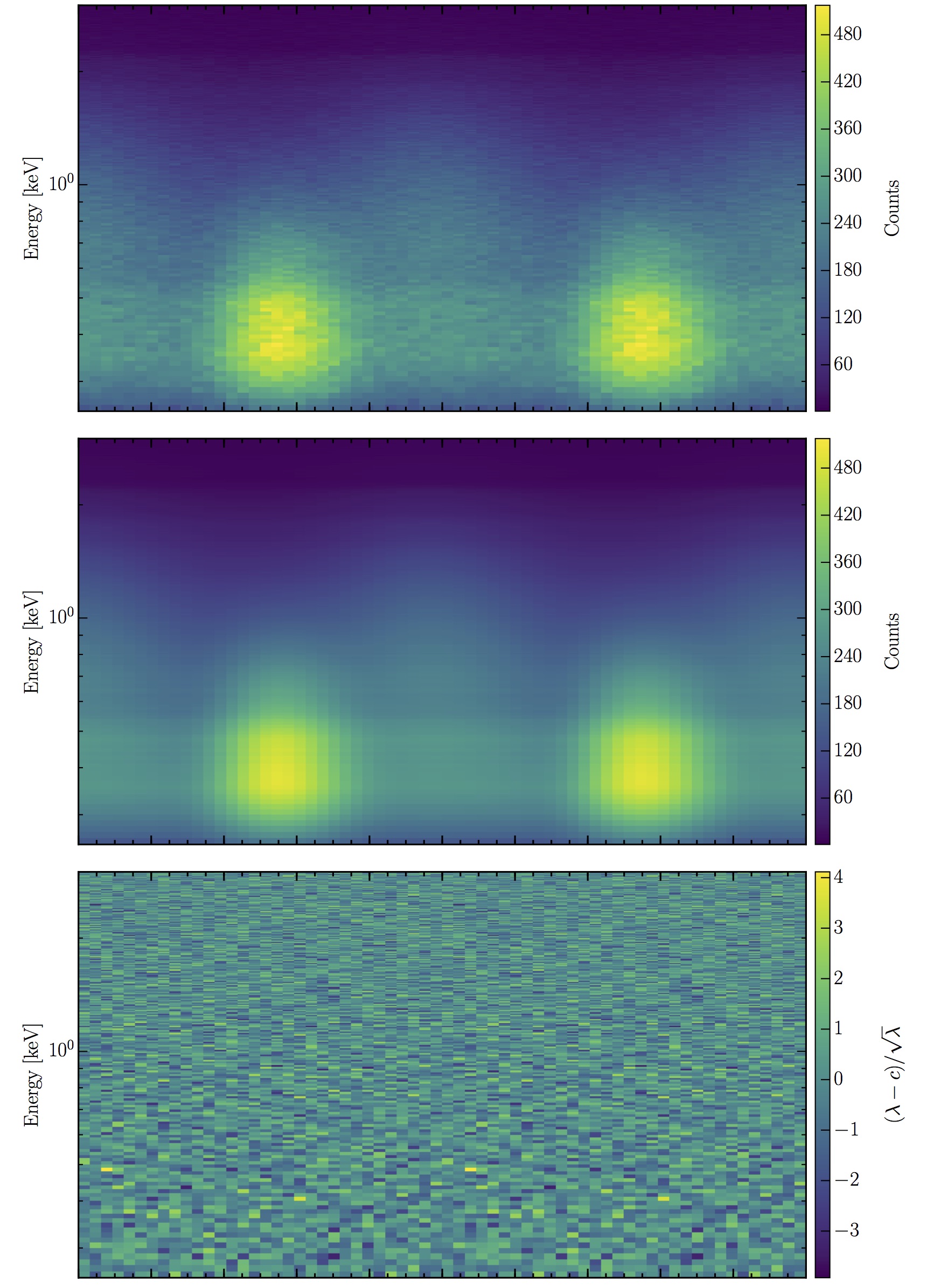}
\includegraphics[width=250pt]{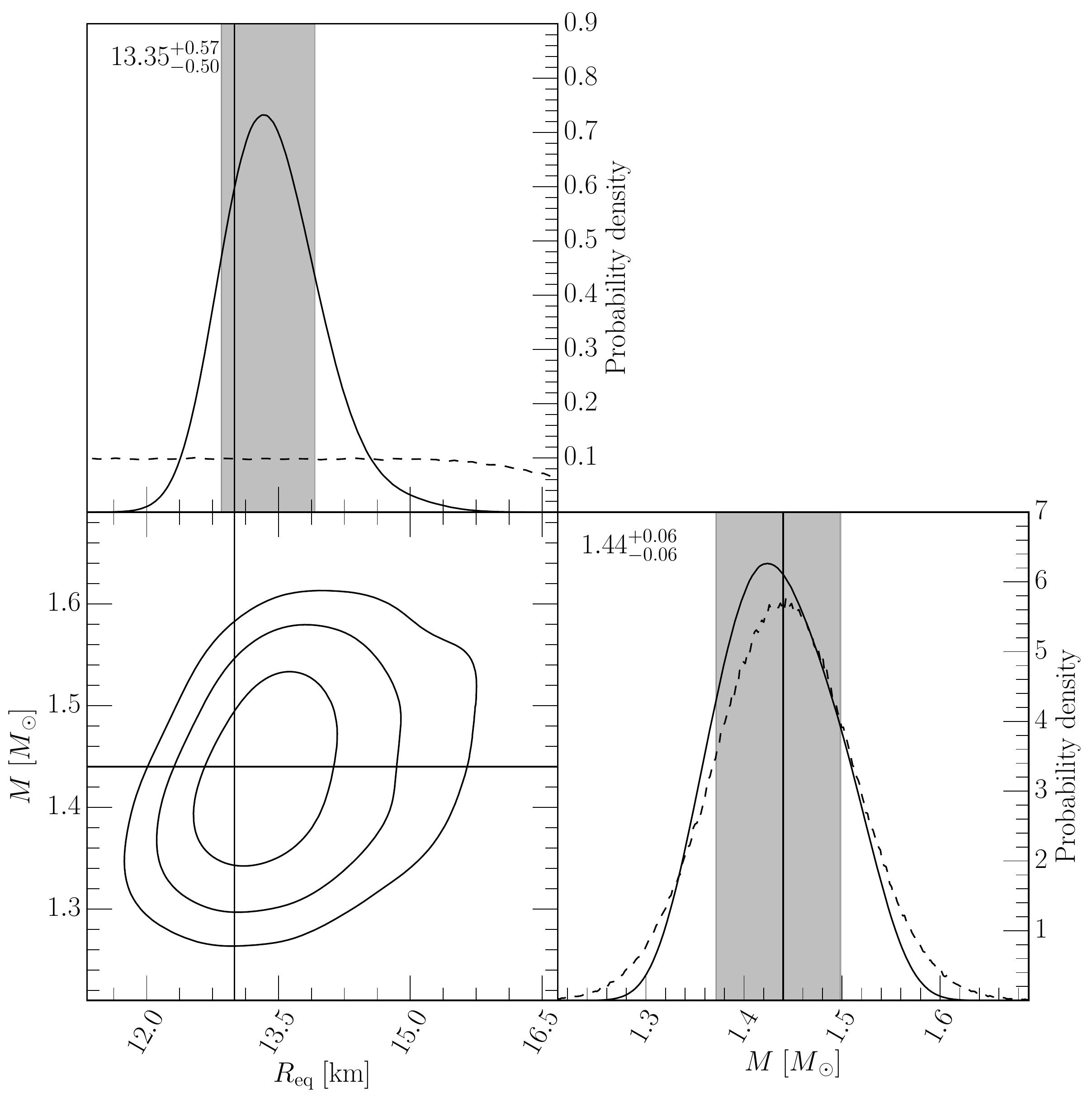}
  \caption{Inference run carried out using the X-PSI code (Riley \& Watts in prep), for synthetic rotation-powered pulsar data similar to that expected from NICER.  The synthetic data were generated with two non-identical single-temperature hotspots and a hydrogen atmosphere model; the inference assumes the same model. Left color panels: The top panel shows phase-energy channel synthetic count data. The center panel shows the result of the inference: the posterior-mean phase-energy interval count pulse from the star, added to the inferred background. The lower panel shows residuals, used for visual inspection of systematic model deficiencies (one form of posterior predictive checking). Right: Corner plot showing 2D (1,2,3 $\sigma$ contours) and 1D marginalized posterior distributions for mass M and radius R, with the $\pm 1\sigma$  band shown in grey. Radius is inferred at the $\pm 7$ \% level, mass at $\pm 4$ \% (this example assumes a tight prior on mass such as we might have from radio data, shown as a dashed line in the mass plot).  {\it Figure courtesy of Thomas Riley.}  \label{infeg1}}
\end{figure}

\begin{figure}[!h]
 \centerline{\includegraphics[width=450pt]{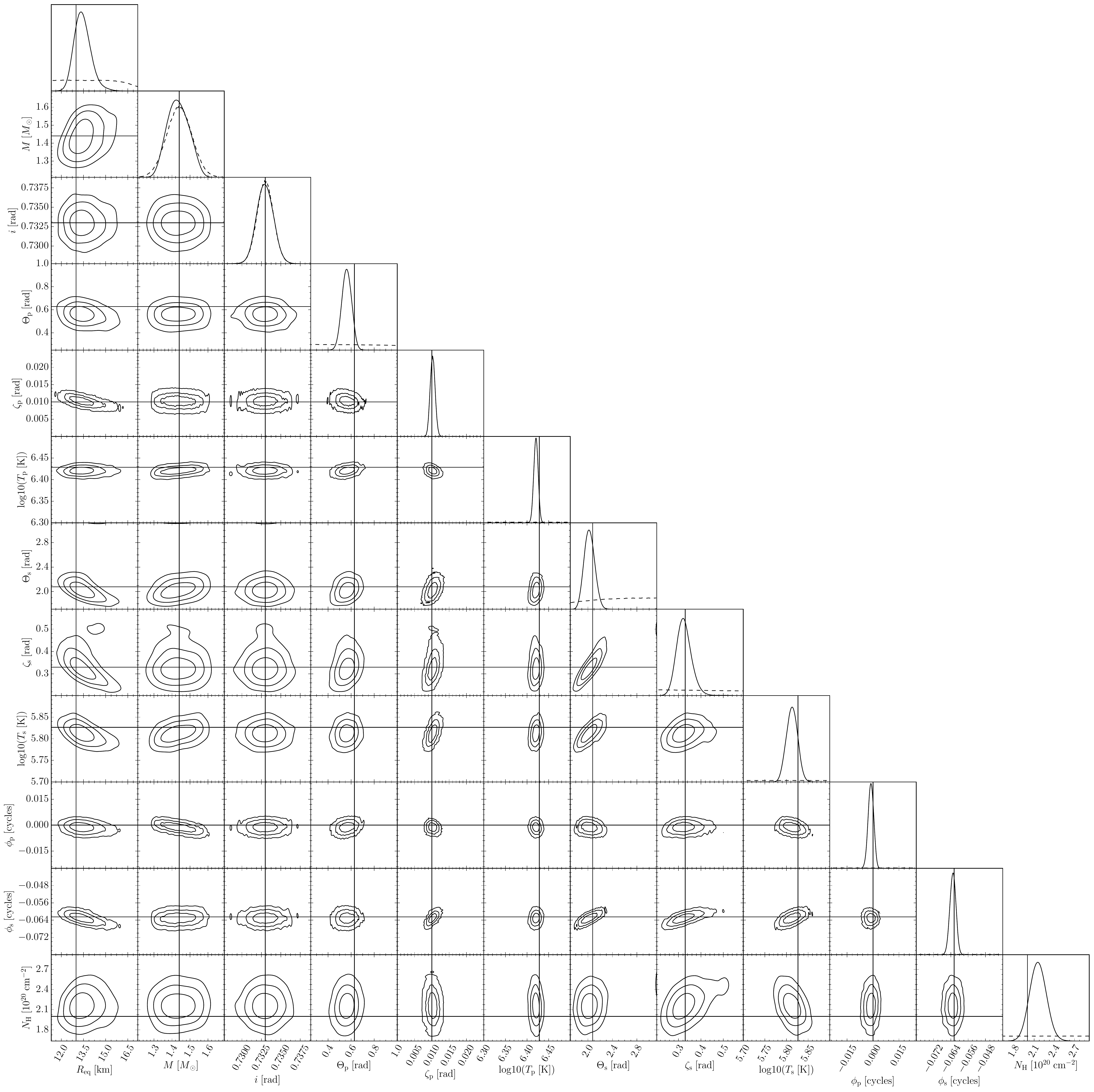}}
  \caption{As Figure \ref{infeg1} but a more extensive corner plot showing 2D (1,2,3 $\sigma$ contours) and 1D marginalized posterior distributions for some of the other parameters that we also infer:  radius R, mass M, observer inclination i, Primary spot centre colatitude, angular radius, temperature T, Secondary spot centre colatitude, angular radius, temperature, spot phases, and hydrogen column density.   Horizontal and vertical solid lines indicate the values used to generate the synthetic data. Dashed lines in the 1D panels indicate the prior: the posteriors are clearly data-dominated. {\it Figure courtesy of Thomas Riley.}  \label{infeg2}}
\end{figure}

NICER has now acquired full data sets for two of the primary rotation-powered pulsars, and assuming nominal operations is expected to acquire a full data set for two more.  Preliminary analyses for the two primary targets indicate that inferring mass and radius posteriors for individual sources at the 5-10\% level (although perhaps with multimodality) is feasible.  NICER will therefore provide the first firm proof of principle that one can do per-source mass-radius inference with rotation-powered pulsars at the few percent level using Pulse Profile Modeling.  Figures \ref{infeg1} and \ref{infeg2} show an example simulation done in support of NICER using synthetic pulsar data, generated using the X-Ray Pulsation Simulation and Inference (X-PSI) code (Riley \& Watts in prep).  X-PSI is a software package for Bayesian modeling of astrophysical X-ray pulsations generated by the rotating, radiating surfaces of relativistic compact stars.  It couples X-ray pulsation likelihood functionality to open-source statistical sampling software like MultiNest \citep{Feroz09} for use on high-performance computing systems.  

\subsection{EQUATION OF STATE CONSTRAINTS}

There are two approaches to EOS inference: one can do direct inference of EOS parameters from the pulse profile data; or one can infer EOS parameters from a set of per-source mass-radius posterior distributions.  The former is computationally as expensive as mass-radius inference, the latter less so. However the latter is something that must be done with care, since there are potential pitfalls, particularly for commonly-used EOS parametrizations, that can bias the results (see \cite{Riley18} and \cite{Raaijmakers18}, which also examine the approaches taken in previous work on EOS inference by \cite{Ozel09,Steiner10,Steiner13,Nattila16,Ozel16,Raithel17}). One method used in the literature to implement the computationally less-expensive approach is to assume that the likelihood function is proportional to the posterior distribution of the exterior space-time parameters (mass and radius); as outlined in \cite{Raaijmakers18} this is an assumption that only holds if the prior on mass and radius is sufficiently non-informative.  Fortunately this is expected to be the case for NICER analysis, even if we use mass priors from radio data during the mass-radius inference, since the original radio analyses use non-informative priors in their computations.  

In \cite{Greif19} we adopted this approach to examine the kind of EOS model parameter constraints that NICER might deliver (assuming two to four idealized mass-radius posteriors of the size anticipated, various different spreads of mass, and two different underlying EOS models).  For the scenarios examined, we found that mass-radius posteriors at the level expected from NICER led to a nominally substantial narrowing of EOS model parameter space (the range of pressures spanned by the 95\% credible band for the EOS at twice nuclear saturation density, for example, was up to a factor $\sim 5$ narrower than the full range considered plausible).  However for some of the scenarios explored, the underlying EOS lay slightly outside the 95\% credible band for some energy densities.  This unreliability in EOS recovery could, we eventually determined, be attributed in part to the EOS model parametrization and choice of priors.    The biases introduced by the EOS models were entirely inadvertent (not deliberately motivated by physics) and had not been noticed in previous studies.  Parameter ranges for the EOS models that we used were picked in order to ensure models spanned a certain range of space, with flat priors on individual parameters; naively this might be considered a conservative choice, and is one that other analyses have also used.  However this choice for the individual EOS model parameters does not correspond to a flat prior in EOS space (pressure as a function of energy density, some function of the individual parameters).  This non-flat EOS prior has a detectable influence when the models are used for EOS inference in a regime where we have only a small number of mass-radius posteriors.  It illustrates the importance of testing inference pipelines using synthetic data; this effect may also be highly relevant for attempts to infer EOS parameters from small numbers of gravitational wave events.  It is not clear if we can resolve this issue with different EOS models.  Precision EOS inference will likely need more or tighter mass-radius posteriors than we can expect from NICER, and will hence require the next generation of telescopes (see Figure 9 of \cite{Ray19} for a simulation done for STROBE-X).

\section{CONCLUSIONS}

The EOS of matter at supranuclear densities remains a major problem in fundamental physics: one that is being studied in the laboratory and via multi-messenger observations of neutron stars. All astronomical techniques to probe the EOS have model dependencies; cross-testing different source types and techniques is imperative. NASA's NICER telescope is poised to deliver the first simultaneous measurements of mass and radius using the Pulse Profile Modeling technique.   We also need to think ahead to the next generation of large area X-ray timing telescopes that will carry out Pulse Profile Modeling for accreting neutron stars; these missions will realize the promise of Pulse Profile Modeling to fully map the dense matter EOS and reveal the nature of subatomic particles at the very highest densities.

\section{ACKNOWLEDGMENTS}
The author would like to thank the organizers of the Xiamen-CUSTIPEN Workshop on the EOS of Dense Neutron-Rich Matter in the Era of Gravitational Wave Astronomy for hospitality and financial support, and acknowledges support from ERC Starting Grant 639217 CSINEUTRONSTAR.

% References

\nocite{*}
\bibliographystyle{aipnum-cp}%
\bibliography{sample}%

\end{document}